\documentclass{PoS}

\title{Redshift measurement of Fermi Blazars for the Cherenkov Telescope Array}

\ShortTitle{Redshift measurement of Fermi Blazars for the Cherenkov Telescope Array}

\author{
\speaker{P. Goldoni}$^1$, S. Pita$^1$, C. Boisson$^2$,
G. Cotter$^3$, D.A. Williams$^4$ and E. Lindfors$^5$ for the CTA
Consortium\footnote{Full consortium author list at http://cta-observatory.org}\\
\llap{$^1$}APC, Univ Paris Diderot, CNRS/IN2P3, CEA/Irfu, Obs de
Paris, Sorbonne Paris Cit\'e, France, 10 rue Alice Domon et Leonie
Duquet, 75205, Paris Cedex 13, France\\
\llap{$^2$} Observatoire de Paris, LUTH, CNRS, Universit{\'e} Paris Diderot, PSL, France\\
\llap{$^3$} Department of Astrophysics, University of Oxford OX1 3RH, UK\\
\llap{$^4$} University of California, Santa Cruz, CA 95064, USA\\
\llap{$^5$} Tuorla Observatory, University of Turku, Finland\\
E-mail: \email{goldoni@apc.univ-paris7.fr},
\email{pita@apc.univ-paris7.fr},
\email{catherine.boisson@obspm.fr},
\email{Garret.Cotter@physics.ox.ac.uk},
\email{daw@ucsc.edu},
\email{elilin@utu.fi}
} 

\abstract{Blazars are active galactic nuclei, and the most numerous
 High Energy (HE) and Very High Energy (VHE) $\gamma$-ray emitters. Their
optical emission is often dominated by non-thermal, and, in the case of BL Lacs,
featureless continuum radiation. This renders the determination of their
redshift extremely difficult. Indeed, as of today only about 50\% of $\gamma$-ray
blazars have a measured spectroscopic redshift. The knowledge of redshift is
fundamental because it allows the precise modeling of the VHE emission
and also of its interaction with the extragalactic background light (EBL).
The beginning of the Cherenkov Telescope Array (CTA) operations in the near future
will allow the detection of several hundreds of new BL Lacs. Using the first Fermi
catalogue of sources above 10 GeV (1FHL), we performed simulations which
demonstrate that at least half of the 1FHL BL Lacs detectable by CTA will
not have a measured redshift. Indeed the organization of observing campaigns to
measure the redshift of these blazars has been recognized as a necessary support 
for the AGN Key Science Project of CTA.
Taking advantage of the recent success of an X-shooter GTO observing campaign, we
thus devised an observing campaign to measure the redshifts of as many as possible
of these candidates. The main characteristic of this campaign with respect to previous
ones will be the use of higher resolution spectrographs and of 8 meter class telescopes.
We are starting submitting proposals for these observations. In this paper we will briefly
describe the selection of the candidates, the characteristics of these observation
and the expected results.}

\FullConference{The 34th International Cosmic Ray Conference,\\
		30 July- 6 August, 2015\\
		The Hague, The Netherlands}

\begin{document}

\section{Introduction}

Blazars are a type of active galactic nuclei possessing relativistic
plasma jets pointing toward the observer (see \cite{Urry1995} for a
review). They are the most numerous $\gamma$-ray emitters in the MeV
to TeV range. The majority of the energy emitted comes from the
radiation of particles accelerated in these jets, which is further
Doppler boosted along the line of sight. This tends to hide the
spectral signature of the host galaxy and renders, especially for BL
lacs objects, the determination of the redshift difficult.
Their spectral energy distribution is characterized by two bumps,
which are generally explained in the framework of one-zone leptonic
models referred to as Synchrotron self-Compton (SSC) )\cite{Ginz1965},
\cite{Ghise1989}).
Alternatively the second peak may be produced
by hadronic processes often associated to pion decay
(e.g. \cite{Mann1993}). In this case BL Lacs may be a 
source of high energy cosmic rays and/or neutrinos \cite{Dermer2001}.
It is also possible that the radiation is produced at great distance
from the BL Lac, if particles from the jet trigger cascades in the
intergalactic space interacting with the CMB or the EBL \cite{Essey2010}.

\begin{figure}
\includegraphics[width=1.\textwidth]{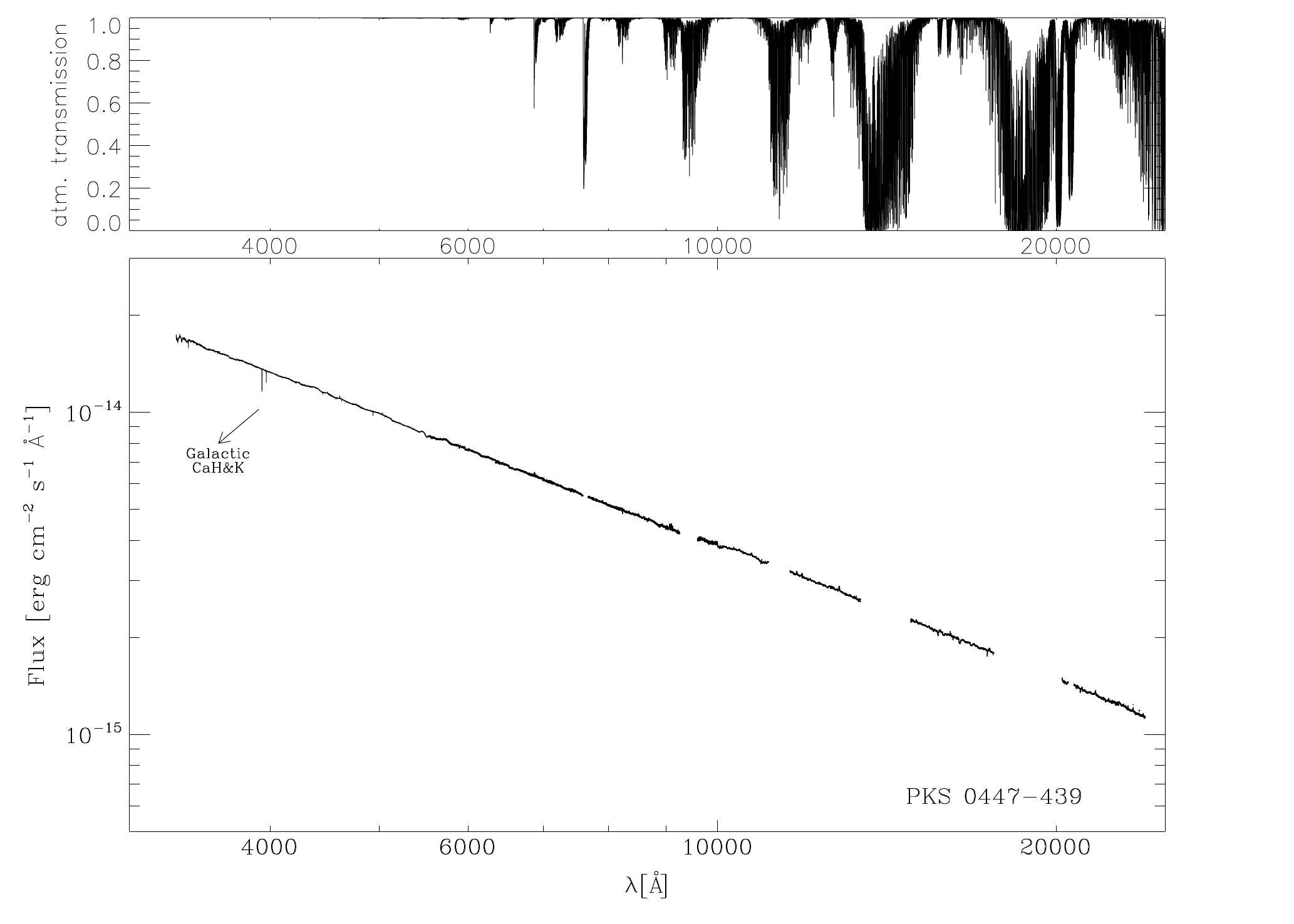}
\caption{UV to NIR spectrum of PKS 0477-439 (from \cite{Pita2014}) obtained with X-shooter,
  the non thermal continuum radiation produces a power law spectrum.
The only spectral feature of astrophysical origin detected in the
spectrum is the Ca H \&K absorption from the interstellar medium of the
Milky Way. In the upper panel we show the
typical Earth atmospheric transmission from Paranal which imprints
telluric absorption features in the spectrum.}
\label{FigPKS}
\end{figure}

At energies greater than 100 GeV or so  VHE $\gamma$-rays are
heavily affected by the interaction with the Extragalactic Background
Light (EBL) \cite{Hauser2001}. The EBL is the
optical to NIR radiation released by the stars and galaxies
throughout the evolution of the Universe. The TeV photons emitted by the
BL Lacs annihilate with the optical ones ($\gamma-\gamma$~interaction)
producing electron-positron pairs. The resulting $\gamma$-ray opacity
is strongly dependent on the distance to the source and dramatically softens
the TeV spectra for energies higher than 1 TeV and redshift $\gtrsim$ 0.2
\cite{Mazin2013}. This effect makes it very difficult to detect
BL Lacs at  ${\rm z} \gtrsim$ 0.5. Conversely, if a complete modeling
of the BL Lac radiation and of the EBL absorption can be achieved,
the EBL properties can be constrained \cite{HESSc2013}.
  
 It is therefore crucial to measure the redshift of a BL Lac. Unfortunately,
the continuum optical/NIR emission of the jet is often much stronger
than the one of the host galaxy. Indeed, BL Lacs were defined as
sources with emission lines with equivalent width (EW) smaller than 5
\AA~\cite{Urry1995}. An example of the featureless continuum of a bright
BL Lac is shown in Figure \ref{FigPKS}. If the host galaxy is not detected, it
is impossible to determine the redshift (and therefore the distance) of the
BL Lac from the spectrum. 
This prevents not only the determination of the amount of line-of-sight
absorption but also the measurement of the luminosity of the source.

   Several deep observing campaigns have been organized in order to
measure BL Lac redshifts, but in each campaign, for a significant
fraction of BL Lacs (e.g. $\sim$ 36 \% in \cite{Pirano2007}), no
spectral feature could be detected. 
After the very extensive recent survey by \cite{Shaw2013}, only $\sim$
44 $\%$ of Fermi BL Lacs have a spectroscopic redshift.
The lack of information on redshift (and therefore luminosity) of 
a great fraction of the sources very strongly biases population studies.
This situation is more disturbing in view of the imminent start of the
operations of CTA.

 As BL Lacs are the dominant source class in the TeV extragalactic sky,
it is expected that CTA will discover hundreds of new BL Lacs, several
of them without redshift. In order to estimate the number of them, we
used the results of the 1FHL Fermi catalog \cite{Ack2013} to simulate
CTA observations. The results show that about half of the newly
discovered VHE BL Lacs will have no measured redshift. A deep search
for their redshift is well justified, but from the above
considerations, the observation method should be chosen carefully.

Recently \cite{Pita2014} have observed a small sample of $\gamma$-ray BL Lacs
using the high spectral resolution spectrograph X-shooter at the ESO
Very Large Telescope (VLT). They have shown that, under these conditions,
a one hour observation is sufficient
to find or constrain the redshift for most of these objects. Their
sample included five objects observed also by \cite{Shaw2013}, who
measured a redshift for only one of them. Conversely \cite{Pita2014} measured three
redshifts and two strict lower limits from line of sight absorption
systems. 
These encouraging results prompted us to organize a program to
measure the redshift of BL Lacs candidates for CTA detection.

 In the following we describe the observations performed by
\cite{Pita2014}, the simulations that allowed us to select
our sample and the plan of the long term observing campaign.

\section{Previous spectroscopic observations }

 In their study of eight HE BL Lacs with unknown or uncertain
 redshift, \cite{Pita2014} determined the redshift of five of them and
constrained the redshift of two others through the detection of
intervening absorbing systems. The observations were performed with
the X-shooter spectrograph \cite{Vernet2011} at the VLT. X-shooter is a high
resolution single object spectrograph sensitive to radiation
from 3000 \AA~ to 24000 \AA. The radiation from the source is split
by two dichroics in three arms, UVB (sensitive from 3000 \AA~ to 5600
\AA), VIS (sensitive from 5500 \AA~ to 10200 \AA) and NIR (sensitive
from 10000 \AA~to 24000 \AA).


\begin{figure}[h!]
\begin{center}
\includegraphics[width=.7\textwidth]{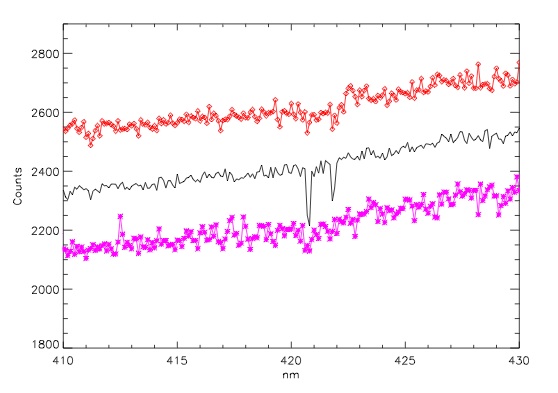}
\end{center}
\caption{We present here the region of the spectrum
  KUV 0033-1198 containing the MgII doublet absorption at z=0.505 as seen in the original (R=10000) X-shooter spectrum \cite{Pita2014}
(center, black line) and the same spectrum convolved to reduce its spectral
resolution to $\sim$ 600 (upper, connected red squares) and $\sim$ 1100 (lower,
connected magenta asterisks). The red and magenta spectrum are artificially offset for display
purposes. It is apparent that identifying the doublet becomes much more difficult
at lower resolution even if the signal to noise ratio is the same.}
\label{mgii_abs}
\end{figure}

 Examination of the results showed that the main factors in this
success were: the high spectral resolution (R=$\lambda \over \Delta \lambda$
$\sim$ 10000), the sensitivity and wavelength range of the instrument
and the good observing conditions. In particular the spectral
resolution is very effective for the detection of narrow spectral
features such as the ones due to intervening absorption systems (Figure
\ref{mgii_abs}). The resolution can also be very helpful for the
correct identification of spectral features produced by the Earth's
atmosphere (see upper box in Figure \ref{FigPKS}).

 The average S/N per pixel reached in these observations was $\sim$ 100 in
the UVB arm, $\sim$ 80 in the VIS arm and $\sim$ 40 in the NIR arm,
the trend of lower S/N with wavelength being mainly due to the spectral shape
(see Figure \ref{FigPKS}). This high S/N combined with the unprecedented
wavelength range permitted enough features to firmly determine the redshift
to be found. Depending on the source, they detected between two and six features,
always including a doublet, see Table \ref{tabLines} for the details.
Moreover, the features that allowed the determination
of the redshift were already detected after one hour of
exposure. Finally, despite reaching a very high S/N, no feature linked
to the host galaxy could be detected in two sources: PKS 0477-439 and KUV 00311-1198. This
suggests that it is not worthwhile to observe for very long exposure
times as for some sources the spectral features cannot be detected
even in these conditions.

\begin{table*}
\caption{\label{tabLines}List of main features detected for each
  source in \cite{Pita2014}. The columns contain: (1) the name of the
  source, (2) the redshift, (3) the name of the absorption and
  emission features detected. Note that Ca II H\&K and Mg II $\lambda$
  2796\,\AA, 2803\,\AA~are doublets, their detection determines
the redshift unambiguously. The majority of these features were detected
after one hour of exposure.}
\centering
\begin{tabular}{lll}
\hline\hline
Source                 & Redshift  & Features \\
\hline
BZB\,J0238--3116   & 0.2329  &  Ca II H\&K, H${\delta}$, Ca I G, Na I
D,  [NII] $\lambda$ 6548\,\AA,  [NII] $\lambda$ 6583\,\AA\\
BZB\,J0543--5532   & 0.273   &  Ca II H\&K, Na I D\\
BZB\,J0505+0415    & 0.424   &  Ca II H\&K, Mgb, CaI+FeI, Na I D\\
RBS\,334           & 0.411   &  Ca II H\&K, Ca I G, CaI+FeI, Na I D, Ca triplet\\
PKS\,0301--243     & 0.2657  &  [OII] $\lambda$ 3727\,\AA, [OIII] $\lambda$ 4959\,\AA, [OIII] $\lambda$ 5007\,\AA\\
                   &         &  [NII] $\lambda$ 6548\,\AA, H${\alpha}$, [NII] $\lambda$ 6583\,\AA\\
\hline
KUV\,00311--1938   & 0.506$\le$z$\le$1.54  & Mg II $\lambda$ 2796\,\AA, 2803\,\AA \\
BZB\,J0816--1311   & 0.288$\le$z$\le$1.56  & Three Mg II $\lambda$ 2796\,\AA, 2803\,\AA\ at different redshifts\\
\hline
\end{tabular}
\end{table*}

\section{Sample Selection}

For our observing program, we selected the most promising candidates for an
early detection by CTA using the recently published 1FHL
catalog  \cite{Ack2013}, which provides a list of blazars
detected by the Fermi-LAT space telescope above 10 GeV
during the three years period from 2008 to 2011.
This selection on the photon energy allows to cull the harder and brighter
sources detected by Fermi. Simple extrapolation of their spectra,
taking into account the effect of EBL absorption (see below for details), show that these sources tend to be good candidates for detection in the VHE domain.

 Out of 514 sources detected at energies higher than 10 GeV, the 1FHL
catalog contains 259 BL Lacs and 58 unidentified AGNs (AGUs), a large fraction
of which could be BL Lacs. The 1FHL catalog also lists a redshift value
(or the lack of it) for all its extragalactic entries. We corrected their lists using the
more recent results of \cite{Pita2014}. After these corrections, 131 BL Lacs and
9 AGUs in the 1FHL catalog have an associated redshift, while 128 BL Lacs
and 49 AGUs do not, 
 
For this project 275 AGNs not yet detected at VHE were selected. Of
these, 110 are BL Lacs and AGUs with redshift (with <$\rm z$> = 0.33), 165 are
BL Lacs and AGUs without redshift.

Taking the source spectral parameters from the 1FHL catalog implies that 
we are using the average spectral parameters of the sources
and therefore these simulations do not consider flaring episodes. 
We also make the hypothesis that the average flux of the selected sources
during the 1FHL catalog integration is representative of their flux in the CTA
era.

 For the sources for which the redshift is unknown, in order to simulate
the EBL absorption, a redshift has to be assumed. For
simplicity we used the same redshift value for all of them.
In the survey of \cite{Shaw2013}, the average redshift is 0.33,
while the average lower limit to the redshift from the detection of
absorption systems is 0.7. Accordingly, we choose three values for our simulations:
0.3, 0.6 and 1. The first choice can be considered slightly optimistic,
the second conservative and the third pessimistic, 
The spectra were then extrapolated into the CTA range applying the
EBL absorption from the models of \cite{Franc2008}.

 For all the sources, with known and unknown redshift, we simulated
 the CTA detection using CTA response functions computed by the consortium.
As these response functions have been determined for zenith angle 20$^{\circ}$,
we put a filter on declination (-55$^{\circ}$,+60$^{\circ}$) considering
$\pm$ 25$^{\circ}$ around the South (Aar) and the North (Tenerife)
sites for which the responses have been computed.
This cuts 17 sources with redshift and 27 sources without redshift from the sample, leaving
us with 93 and 138 sources respectively.

  The exposure time was chosen as to achieve a detection level larger than
5 $\sigma$ and a $\gamma$-ray excess of more than 300 source events.
The results are summarized in Table \ref{tab1} in bins of 10 hours.


\begin{table}[h]
\begin{tabular}{cccccccc}
\hline
 <z> & < 10 h & 10-20 h & 20-30 h & 30-40 h & 40-50 h & > 50 h & Total\\
\hline
{\bf 0.33}  &{\bf 29} & {\bf 6} &{\bf 7} &{\bf 6 }&{\bf 0 }&{\bf 45} &{\bf 93 }\\
\hline
\hline
0.3  & 32 & 23  & 9 & 10 & 2 & 62 & 138 \\
\hline
0.6  & 11 & 11 & 10 & 9 & 8 & 89 & 138 \\
\hline
1   & 1 & 2  & 8  & 1 & 1 & 125 & 138 \\
\hline
\hline
\end{tabular}
\caption{Number of sources detected with > 5 $\sigma$ significance and
more than 300 counts for different CTA exposure times in 10 hours bins. The first line
in boldface lists the results for sources with known redshift. The
other lines list the results for sources with unknown
redshifts for which a redshift of respectively 0.3, 0.6 and 1 have been
assumed.}
\label{tab1}
\end{table}

 From the results in the table it appears that, for <z> $\le$ 0.6
with rather short  observations ($\le$ 20 hours) at least 22 BL Lacs
could be reliably detected. Moreover, it is clear that,
under fair assumptions, a considerable number of
BL Lacs without redshift will be detected, comparable to the numbers
of BL Lacs with known redshift.
 
Details of the simulations will be refined following new results from
Fermi and updated CTA response functions as well as new TeV blazar detection
or publications of new redshift determinations. The list of sources will thus be
continuously evolving in the coming years


\section{Observing campaign definition}

 We have shown that a program to observe $\gamma$-ray BL Lacs optical counterparts
with a high spectral resolution (R$\sim$ 10000) spectrograph at 8 meter class
telescope can consideraby expand the number of $\gamma$-ray BL Lacs with known redshift.
The selected sources are evenly distributed in the Northern and Southern Hemisperes.
However, X-shooter, being installed at VLT, cannot observe them all.
Nonetheless, while X-shooter is a unique instrument due to its NIR arm, similar
instruments in the UVB to VIS range are installed in some of
the World's major telescopes (e.g. ESI at Keck). This survey can therefore be pursued
in the Northern hemisphere by using those instruments.

 Within the CTA consortium, which is a large international collaboration, access to
other facilities is possible and a sustainable observing campaign is being built.
Our  redshift determination collaboration is open to further members who are experienced in this kind of
research and have access to other world class spectrographs similar to
the ones previously mentioned.

 As a start, a subsample of the most promising 80 sources, equally distributed between
Northern and Southern hemispheres  was extracted from our simulations. We are submitting
proposals to ESO to observe with X-shooter at VLT (Southern Hemisphere) and to the W.M. Keck
observatory to observe with ESI at Keck (Northern Hemisphere).
The goal is to observe about 20 sources per year (10 from the Southern hemisphere and
10 from the Northern Hemisphere). However a key objective is to have a sensitivity
as uniform as possible for the whole sample. Therefore, in case of loss of observing time
due to technical problems or bad weather, some sources may be
reobserved at a later time.


\acknowledgments{ We gratefully acknowledge support from the agencies and organizations 
listed under Funding Agencies at this website: http://www.cta-observatory.org/.}




\begin{thebibliography}{99}

\bibitem{Ack2013}
Ackermann, M., Ajello, M., Allafort, A. et al.,
\emph{The First Fermi-LAT Catalog of Sources above 10 GeV},
\emph{ ApJS}
209 34 [{\tt astro-ph/1306.6772}]
\bibitem{Dermer2001} Dermer, C. D. \& Atoyan, A.,
\emph{Prospects for detecting high-energy neutrinos from Blazars},
in proceedings of \emph{27th International Cosmic Ray Conference, 07-15 August, 2001, Hamburg, Germany},
3 1149 [{\tt astro-ph/0107200}]
\bibitem{Essey2010} Essey, W. \& Kusenko, A., 
\emph{A new interpretation of the gamma-ray observations of distant active galactic nuclei},
\emph{Astroparticle Physics}
33 81 [{\tt astro-ph/0905.1162}]
\bibitem{Franc2008}
Franceschini, A., Rodighiero, G., Vaccari, M.,
\emph{Extragalactic optical-infrared background radiation, its time evolution and the cosmic photon-photon opacity},
\emph{A\&A}
487 837 [{\tt astro-ph/0805.1841}]
\bibitem{Ginz1965} Ginzburg, V.L. \& Syrovatskii, S.I., 
\emph{Cosmic Magnetobremsstrahlung (synchrotron Radiation)},
\emph{ARA\&A}
3 297
\bibitem{Ghise1989} Ghisellini, G. \& Maraschi, L.,
\emph{Bulk acceleration in relativistic jets and the spectral properties of blazars},
\emph{ApJ}
340 181
\bibitem{Hauser2001} Hauser, M.G. \& Dwek, E.,
\emph{The Cosmic Infrared Background: Measurements and Implications},
\emph{ARA\&A}
39 249  [{\tt astro-ph/0105539}]
\bibitem{HESSc2013} HESS collaboration, 
\emph{Measurement of the extragalactic background light imprint on the spectra of the brightest blazars observed with H.E.S.S.},
\emph{A\&A}
550 A4 [{\tt astro-ph/1212.3409}]
\bibitem{Mann1993} Mannheim, K.,
\emph{The proton blazar},
\emph{A\&A}
269 67
\bibitem{Mazin2013} Mazin, D., Raue, M., Behera, B. et al.,
\emph{Potential of EBL and cosmology studies with the Cherenkov Telescope Array},
\emph{J. Astroph. Phys}
43 241  [{\tt astro-ph/1303.7124}]
\bibitem{Nolan2012} Nolan, P.L., Abdo, A.A., Ackermann, M. et al.,
\emph{Fermi Large Area Telescope Second Source Catalog},
\emph{ApJS}
199 31  [{\tt astro-ph/1108.1435}]
\bibitem{Pirano2007} Piranomonte, S., Perri, M., Giommi, P. et al.,
\emph{The sedentary survey of extreme high-energy peaked BL Lacs. III. Results from optical spectroscopy},
\emph{A\&A}
470 787 [{\tt astro-ph/0704.1729}]
\bibitem{Pita2014}
Pita, S., Goldoni, P., Boisson, C. et al.,
\emph{Spectroscopy of high-energy BL Lacertae objects with X-shooter
  on the VLT},
\emph{ A\&A}
565 A12 [{\tt astro-ph/1311.3809}]
\bibitem{Shaw2013}
Shaw, M.S., Romani, R.W., Cotter, G. et al.,
\emph{Spectroscopy of the largest ever {\bf $\gamma$}-ray selected BL Lac
  sample}, \emph{ApJ}
764 135 [{\tt astro-ph/1301.0323}]
\bibitem{Urry1995} 
Urry, C.M. \& Padovani, P.,
\emph{Unified Schemes for Radio Loud Active Galactic Nuclei}, \emph{PASP}
{\bf 107}  803  [{\tt astro-ph/9506063}]
\bibitem{Vernet2011} 
Vernet, J., Dekker, H., D'Odorico, S. et al.,
\emph{X-shooter, the new wide band intermediate resolution spectrograph at the ESO Very Large Telescope},
\emph{A\&A}
536, A105  [{\tt astro-ph/1110.1944}]

\end{thebibliography}
\end{document}